\documentclass[aps,prl,preprint,tightenlines,superscriptaddress,
showpacs,byrevtex]{revtex4}
\usepackage{graphicx} 
\usepackage{dcolumn}  
\usepackage{amsmath}
\graphicspath{{ps}}

\newcommand{\mpp}{M_{p\bar{p}}}
\newcommand{\mb}{{M_{\rm bc}}}
\newcommand{\de}{{\Delta{E}}}
\newcommand{\bp}{{B^{+}}}
\newcommand{\bz}{{B^{0}}}
\newcommand{\pp}{{p\bar{p}}}
\newcommand{\ppk}{{p\bar{p}K^+}}

\newcommand{\ppkz}{{p\bar{p}K^0}}
\newcommand{\ppks}{{p\bar{p}K_S^0}}
\newcommand{\ppkst}{{p\bar{p}K^{* +}}}
\newcommand{\ppkstz}{{p\bar{p}K^{*0}}}

\newcommand{\kst}{{K^{*{+}}}}
\newcommand{\kstz}{{K^{*0}}}
\newcommand{\ks}{{K_S^0}}

\begin{document}

\begin{flushright}
                  \noindent \hspace*{3.0in}{
                    Belle Preprint \ \ 2008-5 \\
                    KEK \ Preprint 2007-77
}
\end{flushright}

\title{ \quad\\[0.5cm] Observation of $\bz \to \ppkstz$ 
with a large $K^{*0}$ polarization}

\affiliation{Budker Institute of Nuclear Physics, Novosibirsk}
\affiliation{University of Cincinnati, Cincinnati, Ohio 45221}
\affiliation{Department of Physics, Fu Jen Catholic University, Taipei}
\affiliation{The Graduate University for Advanced Studies, Hayama}
\affiliation{Gyeongsang National University, Chinju}
\affiliation{Hanyang University, Seoul}
\affiliation{University of Hawaii, Honolulu, Hawaii 96822}
\affiliation{High Energy Accelerator Research Organization (KEK), Tsukuba}
\affiliation{Institute of High Energy Physics, Chinese Academy of Sciences, Beijing}
\affiliation{Institute of High Energy Physics, Vienna}
\affiliation{Institute of High Energy Physics, Protvino}
\affiliation{Institute for Theoretical and Experimental Physics, Moscow}
\affiliation{J. Stefan Institute, Ljubljana}
\affiliation{Kanagawa University, Yokohama}
\affiliation{Korea University, Seoul}
\affiliation{Kyungpook National University, Taegu}
\affiliation{\'Ecole Polytechnique F\'ed\'erale de Lausanne (EPFL), Lausanne}
\affiliation{Faculty of Mathematics and Physics, University of Ljubljana, Ljubljana}
\affiliation{University of Maribor, Maribor}
\affiliation{University of Melbourne, School of Physics, Victoria 3010}
\affiliation{Nagoya University, Nagoya}
\affiliation{Nara Women's University, Nara}
\affiliation{National Central University, Chung-li}
\affiliation{National United University, Miao Li}
\affiliation{Department of Physics, National Taiwan University, Taipei}
\affiliation{H. Niewodniczanski Institute of Nuclear Physics, Krakow}
\affiliation{Nippon Dental University, Niigata}
\affiliation{Niigata University, Niigata}
\affiliation{University of Nova Gorica, Nova Gorica}
\affiliation{Osaka City University, Osaka}
\affiliation{Osaka University, Osaka}
\affiliation{Panjab University, Chandigarh}
\affiliation{Saga University, Saga}
\affiliation{University of Science and Technology of China, Hefei}
\affiliation{Seoul National University, Seoul}
\affiliation{Sungkyunkwan University, Suwon}
\affiliation{University of Sydney, Sydney, New South Wales}
\affiliation{Tata Institute of Fundamental Research, Mumbai}
\affiliation{Toho University, Funabashi}
\affiliation{Tohoku Gakuin University, Tagajo}
\affiliation{Tohoku University, Sendai}
\affiliation{Department of Physics, University of Tokyo, Tokyo}
\affiliation{Tokyo Institute of Technology, Tokyo}
\affiliation{Tokyo Metropolitan University, Tokyo}
\affiliation{Tokyo University of Agriculture and Technology, Tokyo}
\affiliation{Virginia Polytechnic Institute and State University, Blacksburg, Virginia 24061}
\affiliation{Yonsei University, Seoul}
\author{J.-H.~Chen}\affiliation{Department of Physics, National Taiwan University, Taipei} 
\author{M.-Z.~Wang}\affiliation{Department of Physics, National Taiwan University, Taipei} 
   \author{I.~Adachi}\affiliation{High Energy Accelerator Research Organization (KEK), Tsukuba} 
   \author{H.~Aihara}\affiliation{Department of Physics, University of Tokyo, Tokyo} 
   \author{K.~Arinstein}\affiliation{Budker Institute of Nuclear Physics, Novosibirsk} 
   \author{V.~Aulchenko}\affiliation{Budker Institute of Nuclear Physics, Novosibirsk} 
   \author{T.~Aushev}\affiliation{\'Ecole Polytechnique F\'ed\'erale de Lausanne (EPFL), Lausanne}\affiliation{Institute for Theoretical and Experimental Physics, Moscow} 
   \author{A.~M.~Bakich}\affiliation{University of Sydney, Sydney, New South Wales} 
   \author{V.~Balagura}\affiliation{Institute for Theoretical and Experimental Physics, Moscow} 
   \author{E.~Barberio}\affiliation{University of Melbourne, School of Physics, Victoria 3010} 
% ...
\collaboration{The Belle Collaboration}

\noaffiliation

\begin{abstract}
Using a $492\,{\rm fb}^{-1}$ data sample
collected near the $\Upsilon(4S)$ resonance
with the Belle detector at the KEKB asymmetric-energy $e^+ e^-$
collider,
we observe the decay $\bz \to \ppkstz$ with a branching fraction 
of $(1.18^{+0.29}_{-0.25} (stat.) \pm 0.11 (syst.)) \times 10^{-6}$.
The statistical significance is $7.2 \sigma$ 
for the signal in the low $p\bar{p}$ mass region.
We study the decay dynamics of $\bz \to \ppkstz$ and compare with 
$\bp \to \ppkst$. 
The $\kstz$ meson is found to be 
almost 100\% polarized (with a fraction of $(101 \pm 13 \pm 3)\%$ in the 
helicity zero state), while the $\kst$ meson has  a $(32 \pm 17 \pm 9)\%$
fraction
in the helicity zero state.
The direct $CP$ asymmetries for $\bz \to \ppkstz$ and $\bp \to \ppkst$ 
are measured to
be $-0.08\pm 0.20\pm 0.02$ and $-0.01\pm 0.19\pm 0.02$, respectively. 
We also study the characteristics of the low mass $p\bar{p}$ enhancements near
threshold and the associated angular distributions. In addition, 
we report improved measurements of the branching fractions   
${\mathcal B}(\bp \to \ppkst) =
(3.38^{+0.73}_{-0.60} \pm 0.39) \times 10^{-6}$ and
${\mathcal B}(B^0 \to \ppkz) = (2.51^{+0.35}_{-0.29} \pm 0.21) \times 10^{-6}$,
which supersede 
our previous measurements.

\end{abstract}

\maketitle

{\renewcommand{\thefootnote}{\fnsymbol{footnote}}
\setcounter{footnote}{0}

After the first observation of the charmless baryonic $B$ meson decay,
$\bp \to \ppk$~\cite{ppk,conjugate}, many three-body charmless 
baryonic decays were
found~\cite{plpi,pph,LLK,plg}. One important and intriguing feature 
of these decays is that the baryon-antibaryon mass distributions all peak near
threshold. However, the BaBar collaboration recently reported
evidence of the decay $\bz \to \ppkstz$ 
but could not establish either the presence or absence of such a threshold
enhancement~\cite{babarnews}. 
On the theoretical side, it is generally 
believed that the $B \to \pp K^*$ decays proceed predominantly 
through a $b \to s$ penguin loop diagram, which could be sensitive to
new physics from heavy virtual particles in the loop.   
Large direct $CP$ violation,
$\sim 20\%$, is predicted 
using 
an effective-amplitude approach in the standard model~\cite{LCP}.
From a pole model~\cite{POLE}, it is expected that 
$\mathcal{B}(\bp \to \ppkst) <
\mathcal{B}(\bp \to \ppk)$ due to the absence of some QCD penguin  
and electroweak 
penguin contributions in the $\ppkst$ mode, and that 
$\mathcal{B}(\bz \to \ppkstz) < \mathcal{B}(\bp \to \ppkst)$
due to the absence of a specific pole contribution 
and the external $W$ emission diagram in the
$\ppkstz$ mode.

In this paper, we study the three-body charmless baryonic decays
$\bz \to \ppkstz (\kstz \to K^+\pi^-)$ and  $\bp \to \ppkst
(\kst \to \ks\pi^+)$. 
The polarization of the $K^*$ meson
is determined, which provides information about the relative 
importance of penguin and external
W-emission contributions~\cite{review}.  
The differential branching fractions as a function of the baryon-antibaryon
mass and  
the polar angle distributions of 
the proton in the baryon-antibaryon system are also presented. 
The direct $CP$ violation parameters of these two decays are also measured.
We use a  492 fb$^{-1}$  data sample,
consisting of 535 $ \times 10^6 B\bar{B}$ pairs,
collected with the Belle detector 
at the KEKB asymmetric-energy $e^+e^-$ (3.5 on 8~GeV) collider~\cite{KEKB}.
The Belle detector is a large-solid-angle magnetic
spectrometer that
consists of a silicon vertex detector (SVD),
a 50-layer central drift chamber (CDC), an array of
aerogel threshold Cherenkov counters (ACC),
a barrel-like arrangement of time-of-flight
scintillation counters (TOF), and an electromagnetic calorimeter
composed of CsI(Tl) crystals located inside
a super-conducting solenoid coil that provides a 1.5~T
magnetic field.  An iron flux-return located outside of
the coil is instrumented to detect $K_L^0$ mesons and to identify
muons.  The detector
is described in detail elsewhere~\cite{Belle}.

The event selection criteria for the primary charged tracks can be found in
Ref.~\cite{Wei}.
$\ks$ candidates are reconstructed as $\pi^+\pi^-$ 
pairs with an invariant mass in the range 
$ 490$ MeV/$c^2 < M_{\pi^+\pi^-} < 510 $ MeV/$c^2$.
The candidate must have a displaced vertex and flight 
direction consistent with
a $\ks$ originating from the interaction point.
We use the selected kaons and pions to form $\kst$
($\to \ks\pi^+$) and  $\kstz$ ($\to K^+\pi^-$) candidates.
Events with a $K^*$ candidate mass between  0.6 GeV/$c^2$ and  1.2 GeV/c$^2$ 
are used for further analysis.
Candidate $B$ mesons are reconstructed in the 
$\bz \to \ppkstz$ and $\bp \to \ppkst$  modes.
We use two kinematic variables in the center-of-mass (CM) frame to identify the
reconstructed $B$ meson candidates: the beam energy
constrained mass $\mb = \sqrt{E^2_{\rm beam}-p^2_B}$, and the
energy difference $\de = E_B - E_{\rm beam}$, where $E_{\rm
beam}$ is the beam energy, and $p_B$ and $E_B$ are the momentum and
energy, respectively, of the reconstructed $B$ meson.
The candidate region is
defined as 5.2 GeV/$c^2 < \mb < 5.3$ GeV/$c^2$ and $-0.1$ GeV $ < \de< 0.3$
GeV. 
The lower bound in $\de$ 
is chosen to exclude possible background from
baryonic $B$ decays with higher multiplicities.
From a GEANT~\cite{geant} based Monte Carlo (MC) simulation, the signal
peaks in a signal box defined by the requirements 
5.27 GeV/$c^2 < \mb < 5.29$ GeV/$c^2$ and $|\de|< 0.05$ GeV.
To ensure the decay process is genuinely charmless, we apply charm vetoes. 
The regions $2.850$ GeV/$c^2 < M_{\pp} < 3.128$ GeV/$c^2$ and 
$3.315$ GeV/$c^2 < M_{\pp} < 3.735$ 
GeV/$c^2$ are excluded to remove background from modes with $\eta_c, J/\psi$ 
and $\psi^{\prime},\chi_{c0},\chi_{c1},h_{c}$ mesons, respectively.
The region $2.262$ GeV/$c^2 < M_{p\ks}, M_{pK^-\pi^+} < 2.310$ GeV/$c^2$ is 
also 
excluded to remove a possible $\Lambda_c^+$ background.  From a study of 
a charmless $B$ decay MC sample, there are non-negligible backgrounds 
in the candidate region
due to $\bp \to \ppk$ and $\bz \to \ppks$. We remove the $B$ candidates 
when their $\mb$ and $\de$ values 
reconstructed for the $\pp K$ hypothesis are in the signal box.

After the above selection cuts,
the background in the fit region arises dominantly from continuum $e^+e^-
\to q\bar{q}$ ($q = u,\ d,\ s,\ c$) processes.
We suppress the jet-like continuum background relative to the more
spherical $B\bar{B}$ signal using a Fisher discriminant~\cite{fisher}
that combines seven event shape variables, as described in Ref.~\cite{etapk}.
Probability density functions (PDFs) for the Fisher discriminant and
the cosine of the angle between the $B$ flight direction
and the beam direction in the $\Upsilon({\rm 4S})$ rest frame
are combined to form the signal (background)
likelihood ${\mathcal L}_{s}$ (${\mathcal L}_{b}$).
The signal PDFs are determined using signal MC
simulation; the background PDFs are obtained from 
the sideband data: 
$5.23$ GeV/$c^2$  $ < \mb < 5.26$ GeV/$c^2$ and  $|\de| < 0.06$ GeV 
for the $\ppkstz$ mode;
$5.25$ GeV/$c^2$  $ < \mb < 5.26$ GeV/$c^2$ and  $|\de| < 0.2$ GeV 
for the $\ppkst$ mode.
The different selections for sideband regions of the two $K^*$ modes 
ensure similar statistics to determine the background PDFs.
We require
the likelihood ratio 
${\mathcal R} = {\mathcal L}_s/({\mathcal L}_s+{\mathcal L}_b)$ 
to be greater than 0.7  for both decay modes.
These selection
criteria are determined by optimization of $n_s/\sqrt{n_s+n_b}$, where $n_s$ 
and $n_b$
denote the expected numbers of signal and background events in the
signal box, respectively. 
We use the branching fractions from our 
previous measurements~\cite{pph} in the calculation of $n_s$ 
and use the number of 
sideband events to estimate $n_b$. 
If there are  multiple $B$ candidates in a single event, we 
select the one with the best $\chi^2$ value from the 
vertex fit.
The fractions of events that have multiple $B$ candidates
are 21\% and 32\% for the $\ppkstz$ and $\ppkst$ modes, respectively.
                               

We perform an unbinned extended
likelihood fit that maximizes the likelihood function
$$ L = {e^{-(n_{K^*}+n_{K\pi}+n_{q\bar{q}})} 
\over N!}\prod_{i=1}^{N}
\left(\mathstrut^{\mathstrut}_{\mathstrut}n_{K^*} P_{K^*}+
n_{K\pi} P_{K\pi}+n_{q\bar{q}} P_{q\bar{q}}\right)$$
to estimate the signal yield of $\pp K^*$ in the region 
$-0.1$ GeV $ < \de< 0.3$ GeV,
5.2 GeV/$c^2 < \mb < 5.3$ GeV/$c^2$ 
and $0.6$ GeV/$c^2 < M_{K\pi} < 1.2$ GeV/$c^2$;
here $N$ is the number of events in the fit, 
and $n_{K^*}, n_{K\pi}$ and $n_{q\bar{q}}$
are fit parameters representing the yields of $B \to p\bar{p}K^*$, 
$B \to p\bar{p}K\pi$ and
continuum background, respectively. 
Each PDF is the product of shapes in
$\mb$, $\de$ and $M_{K\pi}$, which are assumed to 
be uncorrelated,
e.g. for the $i$th event, $P_{p\bar{p}K^*} = 
P_{\mb}(M_{{\rm bc}_i}) \times P_{\de}(\Delta{E}_i) 
\times P_{K\pi}(M_{{K\pi}_i})$.

\begin{figure}[htb]
\hskip -14cm {\bf (a)}\\
\vskip -0.5cm
\includegraphics[width=0.70\textwidth]{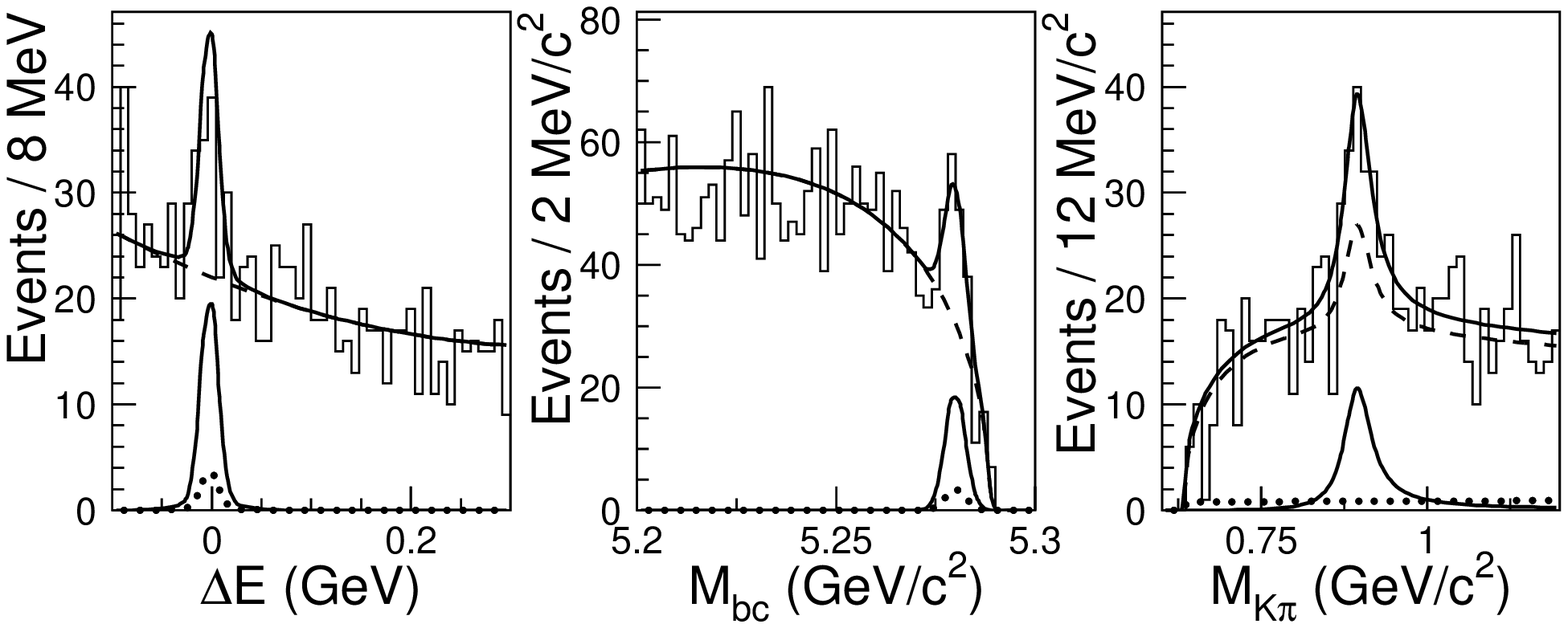}\\
\hskip -14cm {\bf (b)}\\
\vskip -0.5cm
\includegraphics[width=0.70\textwidth]{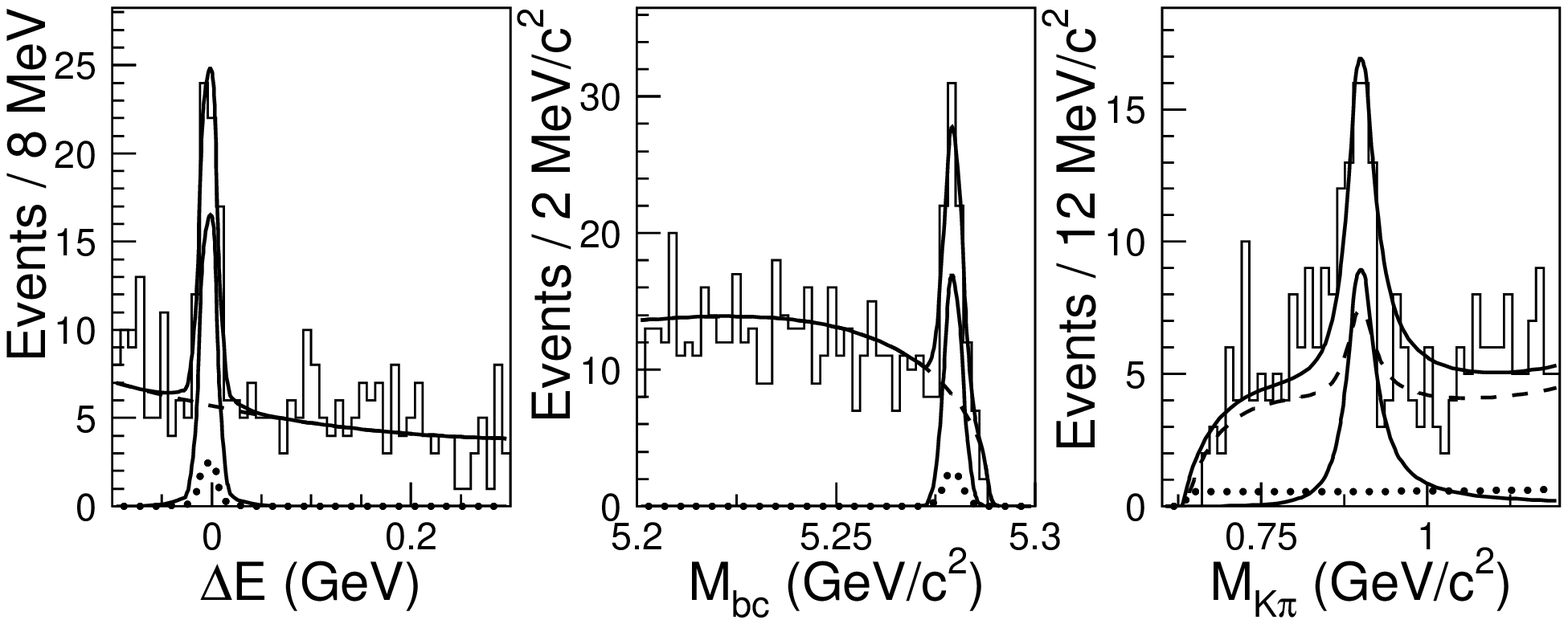}\\
\caption{ Distributions of $\de$ (with $5.27$ GeV/$c^2 < \mb < 5.29$ GeV/$c^2$ and $0.812$ GeV/$c^2 < M_{K\pi} < 0.972$ GeV/$c^2$),
$\mb$ (with $|\de| < 0.05$ GeV and $0.812$ GeV/$c^2 < M_{K\pi} < 0.972$ GeV/$c^2$) 
and $M_{K\pi}$ (with $|\de| < 0.05$ GeV and $5.27$ GeV/$c^2 < \mb < 5.29$ GeV/$c^2$), respectively, 
with proton-antiproton pair mass less than 2.85 GeV/$c^2$ for
(a) $\ppkstz$ and (b) $\ppkst$ modes.
The solid curves, solid peaks, dotted curves and dashed curves represent 
the combined fit
result, fitted $B \to p\bar{p}K^*$ signal, $B \to p\bar{p}K\pi$ signal 
and fitted background, respectively. The areas of dotted curves are about
15\% of those of the solid peaks.}
 
\label{fg:mergembde}

\end{figure}

For the PDFs of $\ppkstz$, $\ppkst$ and $\pp K\pi$ decay modes, 
we use a Gaussian function to represent $P_{\mb}$ and a double Gaussian for $P_{\de}$ 
with parameters determined by MC signal events.
Moreover, we use a p-wave Breit-Wigner function~\cite{Zhang}
 to parameterize the $P_{M_{K\pi}}$  distribution for $\ppkstz$ and $\ppkst$
and use a function obtained by the LASS collaboration~\cite{LASS}
for $\pp K\pi$.
The parameters of these PDFs have been modified to account
for the differences between data and MC using control samples of 
$J/\psi K^{*0}$ and $J/\psi K^{*+}$ with $J/\psi \to \pp$.
The modifications related to the mass peaks are all less than
1 MeV/$c^2$. The $\de$ distribution has a $\sim -3$ MeV shift while the modification for
its width is $\sim 1$ MeV.
For the continuum background PDFs,
we use a parameterization that was first employed by 
the ARGUS collaboration~\cite{Argus}, 
$ f(\mb)\propto \mb\sqrt{1-x^2}
e^{-\xi (1-x^2)}$,  
to model
the $P_{\mb}$ with $x$ given by $\mb/E_{\rm beam}$ and where $\xi$ is
a fit parameter. 
The $P_{\de}$ distribution is modeled by a normalized second-order 
polynomial whose coefficients are fit parameters.
The PDF $P_{M_{K\pi}}$ is modeled by a p-wave function and a threshold function, 
$P_{M_{K\pi}} = r\times P_{p-\textrm{wave}} + (1-r)\times P_{\textrm{threshold}}$ and
$P_{\textrm{threshold}} \propto (M_{K\pi}-M_{K}-M_{\pi})^{s}\times
e^{[c_{1}\times(M_{K\pi}-M_{K}-M_{\pi})+c_{2}\times(M_{K\pi}-M_{K}-M_{\pi})^2]}$
where $r, s, c_{1}$ and $c_{2}$ are fit parameters.
Figure~\ref{fg:mergembde} shows the fits used to obtain the $B \to \pp K^*$ yields
in the proton-antiproton mass region below 2.85 GeV/$c^2$,
which we refer to
as the threshold-mass-enhanced region.
The signal yields are
$70.1^{+14.8}_{-13.9}$ and $54.2^{+10.9}_{-10.1}$ 
with statistical significances of $7.2 $ and $8.8$
standard deviations
for the $\ppkstz$ and $\ppkst$ modes, respectively.   
The significance is defined as $\sqrt{-2{\rm ln}(L_0/L_{\rm max})}$,
where $L_0$ and $L_{\rm max}$ are the likelihood values returned by the 
fit with the signal yield fixed to zero and at its best fit value.

\begin{figure}[htb]
\begin{center}
\hskip -3.5cm {\bf (a)} \hskip 6.35cm {\bf (b)}\\
\vskip -1cm
\includegraphics[width=0.42\textwidth]{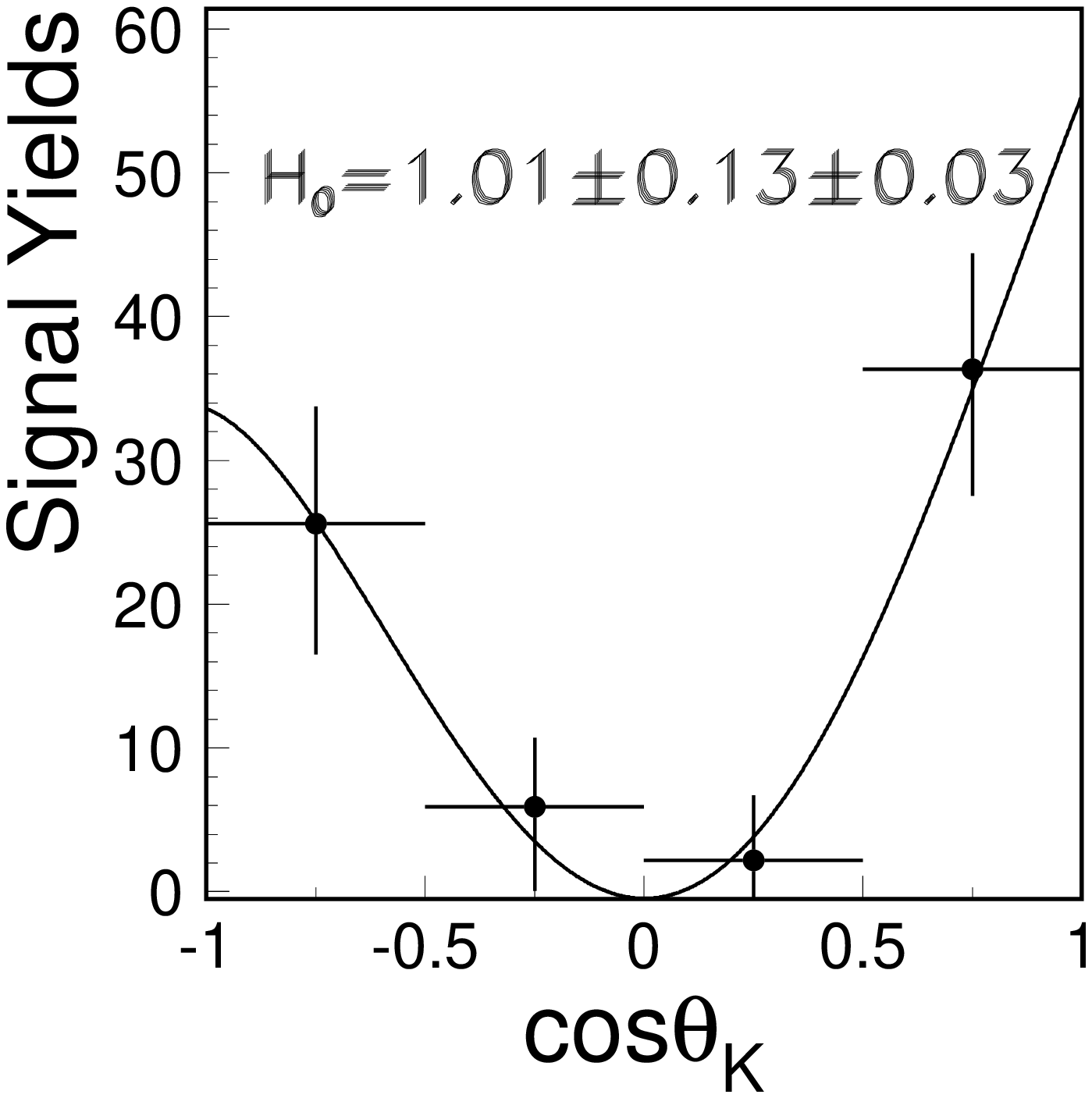}
\includegraphics[width=0.42\textwidth]{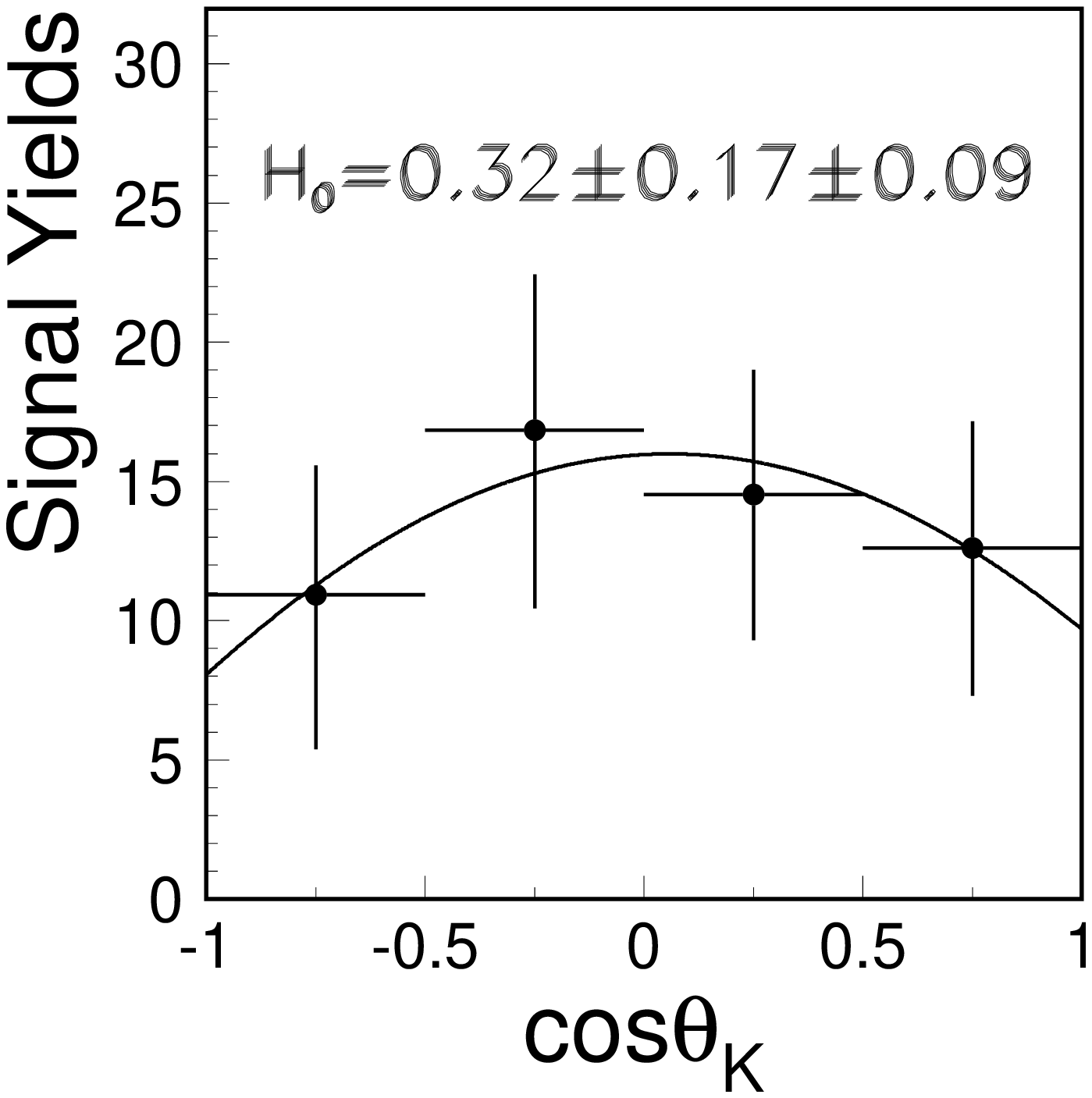}
\end{center}
\caption{
$B$ yield distributions as functions of $\cos\theta_K$ with fit curves 
overlaid for 
(a) the $\ppkstz$ mode and (b) the $\ppkst$ mode.
The fraction of the
signal in the helicity zero state is the fit parameter and is denoted by $H_0$.
The asymmetries in the fit curves are due to detection efficiencies.
The underlying theoretical distributions are symmetric.
}
\label{fg:helicity}
\end{figure}

We determine the angular distribution of the $K^*$ meson 
in the region $M_{\pp} < 2.85$ GeV/$c^2$
using likelihood fits to 
obtain signal yields in bins of $\cos\theta_K$, where $\theta_K$ is 
the polar angle of the $K$ meson in the $K^*$ helicity frame.  
The theoretical PDF for the $K^*$ meson has the form 
3/2 $\cos^2\theta_K$ for a pure helicity zero 
state and 3/4 $\sin^2\theta_K$ for a pure helicity one ($\pm 1$) state. 
We use MC simulation to obtain the efficiency function and convolve it with
the theoretical forms in order to obtain the final PDFs for different helicity 
states.
The signal yields in bins of $\cos\theta_K$ are then fitted with
the above two different PDFs where the fraction of the helicity zero state
is floated in the fit and the total yield is fixed to the experimental result.
The $B$ yield distributions in bins of $\cos \theta_K$ 
with the corresponding fit curves are shown 
in Fig \ref{fg:helicity}. We find that the $\kstz$ meson has a fraction of
$(101 \pm 13 \pm 3)\%$ in the
helicity zero state and the $\kst$ meson has  a $(32 \pm 17 \pm 9)\%$ fraction
in the helicity zero state. It is interesting to note that the helicity 
zero amplitude is expected to be dominant in the $b \to s$ penguin transition
due to the ($V-A$) nature of the weak interaction and 
helicity conservation in the strong interaction~\cite{review}.
The systematic uncertainty is obtained from the 
$B \to J/\psi K^*, J/\psi \to \mu^+ \mu^-$ control sample. We compare our
measured $K^*$ polarization in the helicity zero state
with the PDG value~\cite{PDG}. The difference is added in quadrature 
with the PDG error and the fit error to extract the final systematic 
uncertainty. These uncertainties are 0.03 and 0.09 for the $\ppkstz$ and $\ppkst$ modes, 
respectively.       
\begin{figure}[htb]
\vskip 1.5cm
\begin{center}
\hskip -3.5cm {\bf (a)} \hskip 6.52cm {\bf (b)}\\
\vskip -1cm
\includegraphics[width=0.43\textwidth]{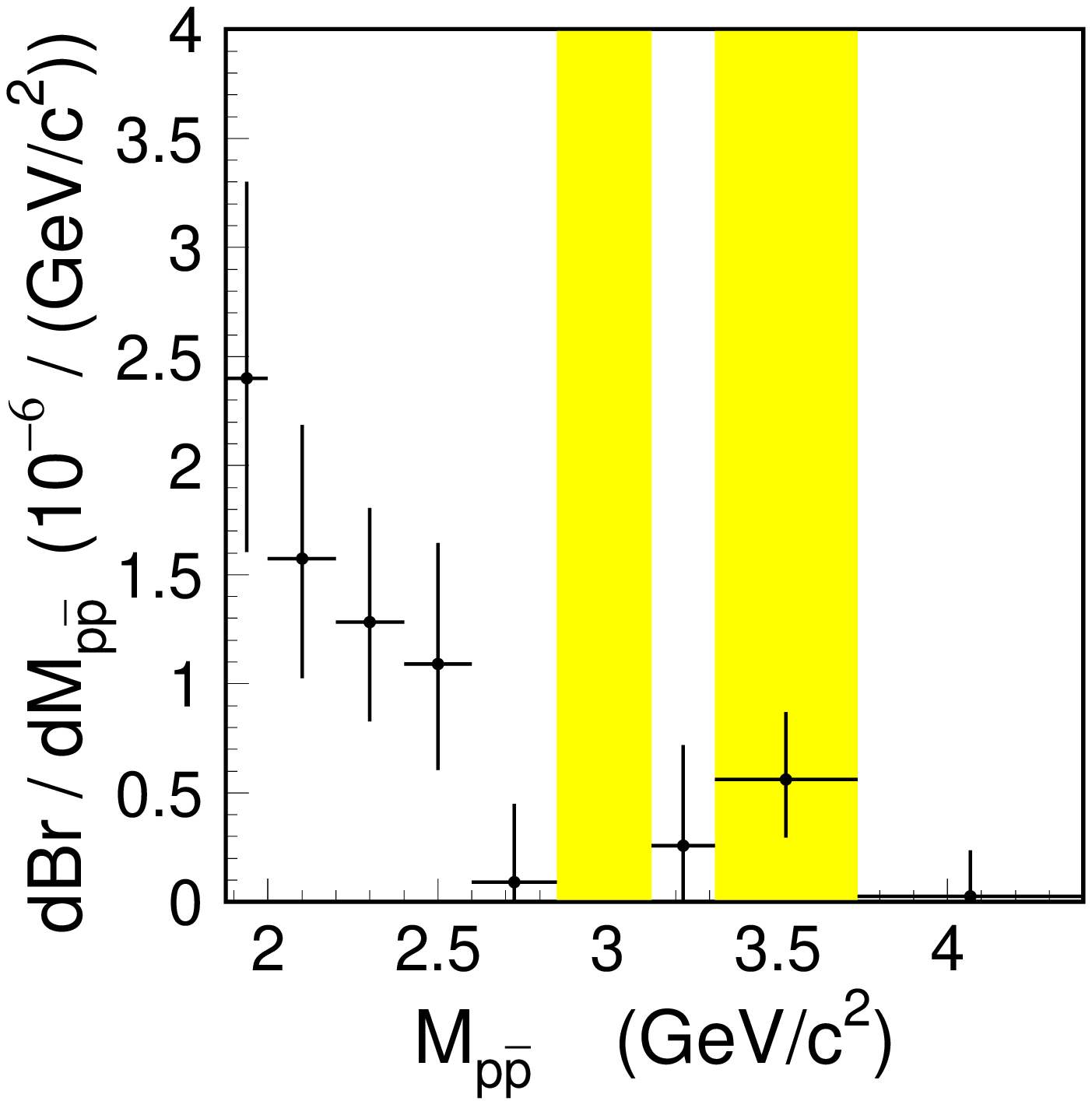}
\includegraphics[width=0.43\textwidth]{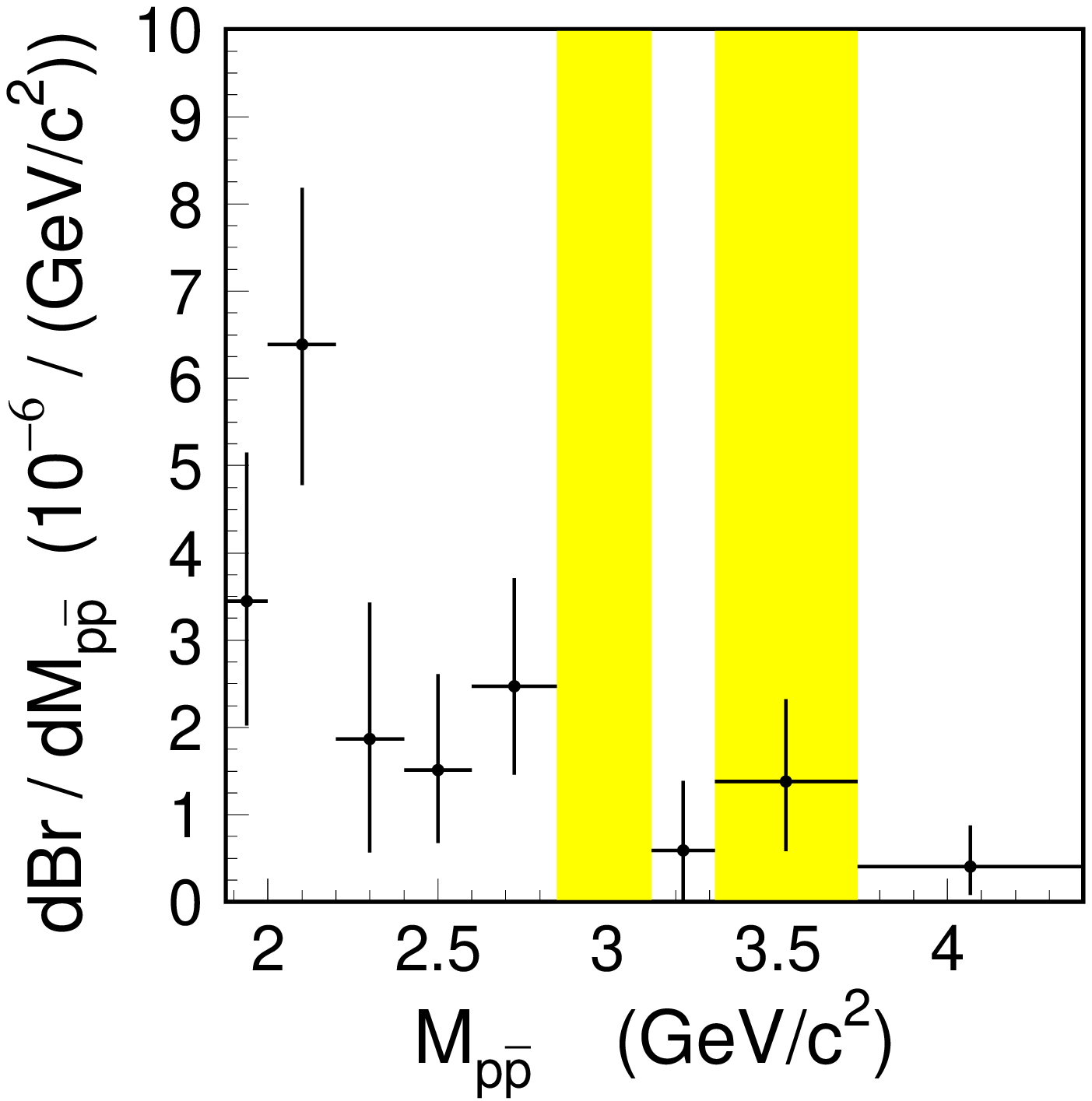}
\caption{Differential branching fractions for
(a) the $\ppkstz$ and (b) the $\ppkst$
modes 
as a function of proton-antiproton invariant mass.
Note that the two shaded mass bins contain charmonium events and
are excluded from the charmless signal yields.
The data points for the $2.85$ GeV/$c^2 <M_{p \bar{p}}<3.128$ GeV/$c^2$
mass region are off-scale.
}
\label{fg:allphase}
\end{center}
\end{figure}

Since the detection efficiency depends on $\mpp$,
we determine the $B \to \pp K^*$ yields in bins of $\mpp$.
We generate a large phase-space MC sample in order
to estimate the efficiencies properly where the sub-decay branching
fractions of $K^*$ to corresponding final states
are included. The $K\pi$ angular distribution
is fixed by the measured $K^*$ polarization for all $\mpp$ bins.
The partial branching fractions are obtained by correcting
the fitted $B$ yields for the mass-dependent efficiencies.
The differential branching fractions as a function of the
proton-antiproton  mass for both $\ppkstz$ and $\ppkst$ modes
are shown in Fig.~\ref{fg:allphase}, and the measured branching
fractions for different $\mpp$ bins are listed in
Table~\ref{ppkst-bins}. 
Applying 3.075~GeV/$c^2 < M_{\pp} < 3.117$~GeV/$c^2$ for $J/\psi$ selection,
we find good agreement, within $1 \sigma$,
between our branching fraction measurements and 
the PDG values~\cite{PDG}.  
In contrast to Ref.~\cite{babarnews}, 
we find that a threshold enhancement
is present for the $\bz \to \ppkstz$ decay. 
With the charmonium regions excluded, 
we sum these partial branching fractions 
to obtain:
${\mathcal B}(\bz \to \ppkstz) = (1.18^{+0.29}_{-0.25}
\pm 0.11 )\times 10^{-6}$ and ${\mathcal B}(\bp \to \ppkst) =
(3.38^{+0.73}_{-0.60} \pm 0.39) \times 10^{-6}$.
As a by-product of our analysis, 
we also use the $\bp \to \ppk$ and $\bz \to \ppks$ signals to estimate 
the corresponding branching fractions in different $\mpp$ bins. 
The total charmless branching fraction
$\mathcal{B} (\bp \to \ppk)$ is $(5.36^{+0.23}_{-0.22})
\times 10^{-6}$, which agrees well with our latest
results, $(5.54^{+0.27}_{-0.25}
\pm 0.36 )\times 10^{-6}$~\cite{Wei}.
The measured value of ${\mathcal B}(B^0 \to \ppkz)$ is
$(2.51^{+0.35}_{-0.29} \pm 0.21) \times 10^{-6}$. 
This result also supersedes our previous measurement~\cite{pph}.
With improved experimental accuracy, the following relationships
$\mathcal{B}(\bp \to \ppk) > \mathcal{B}(\bp \to \ppkst)$ and 
$\mathcal{B}(\bp \to \ppkst) > \mathcal{B}(\bz \to \ppkstz)$ are established.
These inequalities agree with the pole model predictions~\cite{POLE}, but
the measured
$\mathcal{B}(\bz \to \ppkstz)$ is about a factor of 20 larger than 
predicted. This may indicate that the relative weights
of different pole contributions in Ref.~\cite{POLE} are incorrect.

\begin{table}[htb]
\caption{Signal yields and branching fractions ${\cal B}$ ($10^{-6}$) 
in different $\mpp$ regions for $\bz\to\ppkstz$(left) and $\bp\to\ppkst$(right).}
\begin{center}
\begin{tabular}{@{}l|rc|rc@{}}
 & \multicolumn{2}{c|}{$\ppkstz$} & \multicolumn{2}{c}{$\ppkst$} \\
$\mpp$ (GeV)& \multicolumn{1}{c}{Yield} &${\cal B}$ ($10^{-6}$)
            & \multicolumn{1}{c}{Yield} &${\cal B}$ ($10^{-6}$)\\
\hline
$<2.0$& $21.4^{+8.0}_{-7.1}$& $0.30^{+0.11}_{-0.10}$ 
      & $9.0^{+4.4}_{-3.7}$ &  $0.43^{+0.21}_{-0.18}$\\
$2.0-2.2$& $21.5^{+8.4}_{-7.5}$& $0.31^{+0.12}_{-0.11}$
         & $25.1^{+7.1}_{-6.3}$ & $1.28^{+0.36}_{-0.32}$\\
$2.2-2.4$& $15.7^{+6.4}_{-5.6}$ & $0.26^{+0.10}_{-0.09}$
         & $6.4^{+5.4}_{-4.5}$  & $0.37^{+0.31}_{-0.26}$\\
$2.4-2.6$& $12.3^{+6.2}_{-5.4}$ & $0.22^{+0.11}_{-0.10}$
         & $4.5^{+3.3}_{-2.5}$ & $0.30^{+0.22}_{-0.17}$\\
$2.6-2.85$& $1.2^{+4.9}_{-3.9}$ & $0.02^{+0.09}_{-0.07}$
          & $9.6^{+4.8}_{-3.9}$ & $0.62^{+0.31}_{-0.25}$\\
$2.85-3.128$(veto)& $224.2^{+18.2}_{-17.6}$ & $4.12^{+0.34}_{-0.32}$
                  & $55.7^{+9.8}_{-9.0}$  & $3.66^{+0.65}_{-0.59}$\\
$3.128-3.315$& $2.6^{+4.7}_{-3.5}$ & $0.05^{+0.09}_{-0.06}$
             & $1.5^{+2.1}_{-1.5}$ & $0.11^{+0.15}_{-0.11}$\\
$3.315-3.735$(veto)& $11.9^{+6.6}_{-5.6}$ & $0.24^{+0.13}_{-0.11}$
                   & $7.1^{+4.8}_{-4.1}$ & $0.58^{+0.40}_{-0.34}$\\
$>3.735$& $0.7^{+5.5}_{-4.4}$ & $0.02^{+0.14}_{-0.11}$
           & $2.5^{+2.9}_{-2.0}$ & $0.27^{+0.31}_{-0.22}$\\
\hline 
Charmless & $75.4^{+17.1}_{-14.7}$ & $1.18^{+0.29}_{-0.25}$
               & $58.7^{+12.1}_{-10.1}$ & $3.38^{+0.73}_{-0.60}$\\
\end{tabular}
\end{center}
\label{ppkst-bins}
\end{table}

Systematic uncertainties 
are determined using high-statistics control data samples. For proton
identification, we use a  $\Lambda \to p \pi^-$ sample, while for
$K/\pi$ identification we use a $D^{*+} \to D^0\pi^+$,
 $D^0 \to K^-\pi^+$ sample.
Note that the average
efficiency difference for PID
between data and MC has been corrected to obtain the final branching fraction
measurements. The corrections are about 11.5\% and 11.7\% for the $\ppkstz$
and $\ppkst$ modes, respectively. 
The uncertainties associated with
the PID corrections are estimated to be 4\% for two protons
and 1\% for one kaon/pion.
The tracking uncertainty is determined with
fully and partially reconstructed $D^*$ samples. It is about 1\% per
charged track.
The uncertainty in $\ks$ reconstruction
is determined to be 4\% from a sample of $D^- \to \ks\pi^-$ events.
The $\mathcal R$ continuum suppression uncertainty of 2.3\% 
is estimated from control samples with similar final states,
$B \to J/\psi K^*$  with $J/\psi \to \mu^+\mu^-$.
The uncertainties in the best $B$ candidate selection are estimated
to be 2.0\% and 3.5\% for the $\ppkstz$ and $\ppkst$ modes, respectively, 
by taking a difference in the branching fractions with and without 
the best candidate selection.
A systematic uncertainty of 5.2\% in the fit yield is determined by varying
the parameters 
(or changing the functional forms) of the signal and background PDFs.  
The MC statistical
uncertainty 
is less than 3\%. 
The efficiency error caused by the $K^*$ polarization modeling is
estimated to be 2.4\% and 4.0\% for the $\ppkstz$ and $\ppkst$ modes, 
respectively, by
changing the polarization value by $\pm 1 \sigma$.
The error on the
number of $B\bar{B}$ pairs is 1.3\%, where we
assume that the branching fractions of $\Upsilon({\rm 4S})$
to neutral and charged $B\bar{B}$ pairs are equal.
We first sum the correlated errors linearly and then combine them with the
uncorrelated ones in quadrature. The total systematic
uncertainties are 9.7\% and  11.6\% for
the $\ppkstz$ and $\ppkst$ modes, respectively.

We study the proton angular distribution in the proton-antiproton
helicity frame with $M_\pp < 2.85$ GeV/$c^2$.
The angle $\theta_p$ is defined as the
angle between the proton direction and the positive strangeness $K^*$ 
(i.e. $K^{*-}$ or $\bar{K}^{*0}$) direction
in the proton-antiproton pair rest frame. The $\cos \theta_p$ distributions, 
shown in Fig.~\ref{fg:ppkst-costhp}, do not have a  prominent peaking 
feature toward $\cos \theta_p \sim 1$, which was first observed in 
the decay $\bp \to \ppk$~\cite{polar}. 
However, current statistics are inadequate to draw any definitive conclusions
about $B \to p\bar{p}K^*$.

\begin{figure}[htb]
\vskip 1.5cm
\begin{center}
\hskip -3.2cm {\bf (a)} \hskip 6.62cm {\bf (b)}\\
\vskip -1cm
\includegraphics[width=0.43\textwidth]{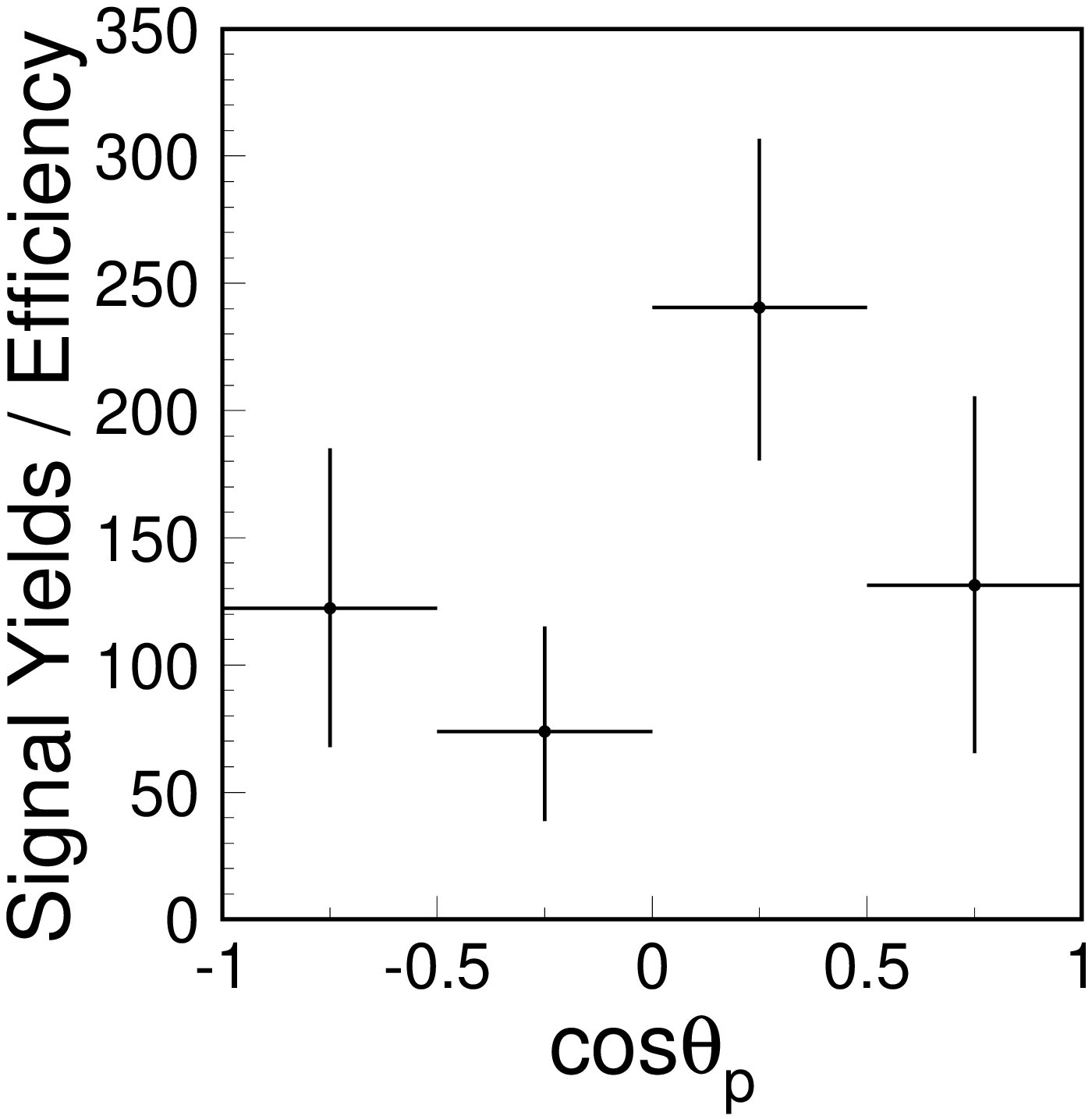}
\includegraphics[width=0.43\textwidth]{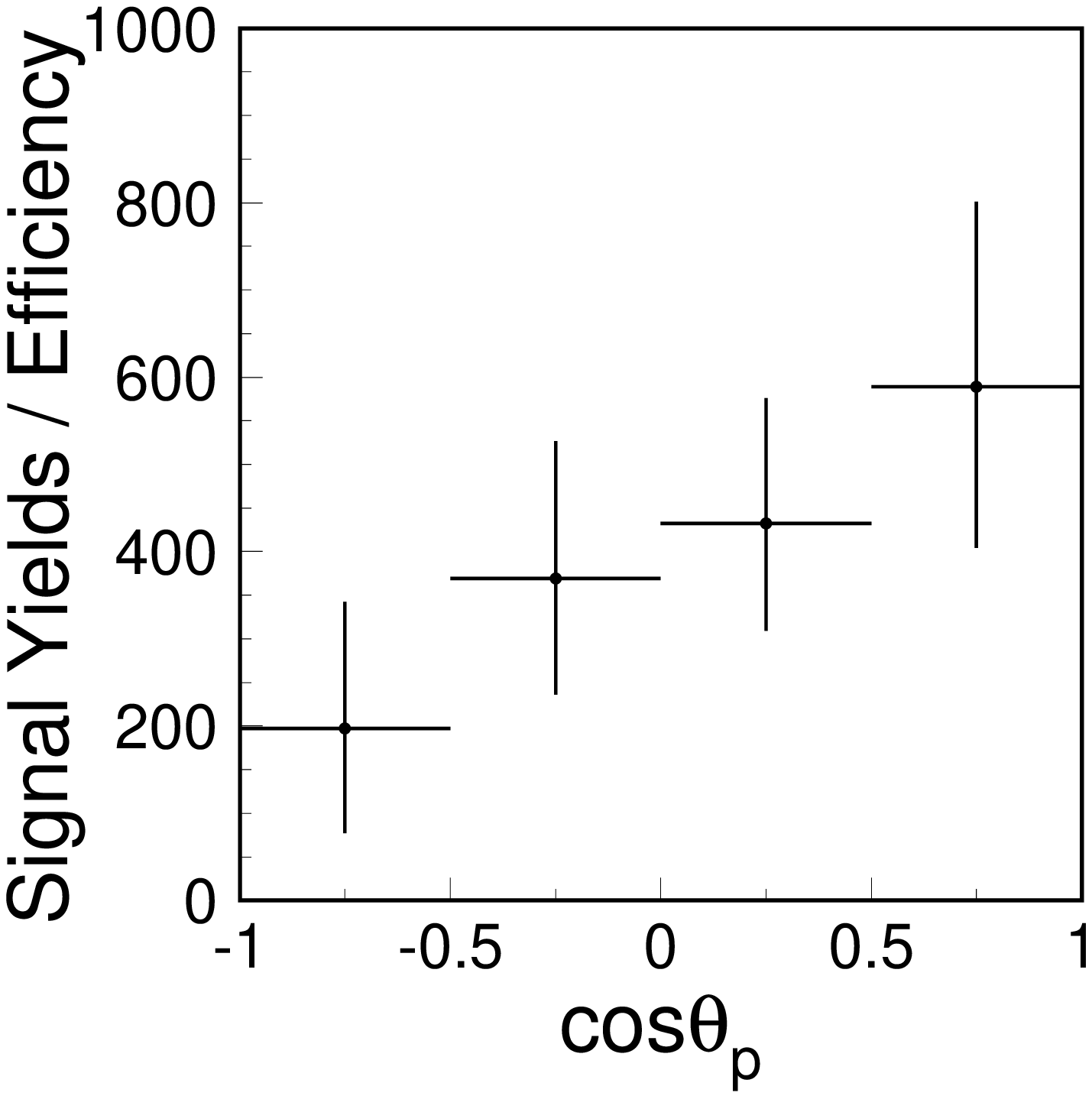}
\caption{Distributions of efficiency corrected signal yields
{\it vs.} $\cos \theta_p$ 
 in the proton-antiproton system 
 with $M_{\pp} < 2.85$ GeV/$c^2$ for (a) $\bz \to \ppkstz$ and
(b) $\bp \to \ppkst$.
}
\label{fg:ppkst-costhp}
\end{center}
\end{figure}

To examine the prediction~\cite{LCP} that direct $CP$ violation
in $\bp \to \ppkst$ can be as large as $\sim 20\%$, 
we define the charge asymmetry $A_{ch}$
as $(N_{b} - N_{\bar{b}})/ (N_{b} + N_{\bar{b}})$ for the $\ppkstz$
and $\ppkst$ modes,
where 
$N$ and $b$ stand for
the efficiency corrected $B$ yield and quark flavor, respectively. 
The results are $-0.08\pm 0.20\pm 0.02$ and $-0.01\pm 0.19\pm 0.02$ 
for the $\ppkstz$ and $\ppkst$ modes, respectively.
The systematic uncertainty is estimated from the measured charge 
asymmetry for the sideband data.

In summary, using 535 $ \times 10^6 B\bar{B}$ events and applying
charmonium vetoes, we 
observe the $\bz \to \ppkstz$ decay with a branching fraction
of $(1.18^{+0.29}_{-0.25} (stat.) \pm 0.11 (syst.)) \times 10^{-6}$.
The signal yield is $70.1^{+14.8}_{-13.9}$ with a significance of 
7.2 standard deviations in the $\mpp<2.85$ GeV/$c^2$ mass region.
The $\kstz$ meson is found to be
$(101 \pm 13 \pm 3)\%$ in the
helicity zero state, compared to 
$(32 \pm 17 \pm 9)\%$ for the $\kst$ meson.
The smaller $K^{*+}$ polarization in the $\ppkst$ decay may be attributed
to an additional contribution from external $W$ emission.
We also observe a low mass $p\bar{p}$ enhancement near
threshold for the $\ppkstz$ mode.  
The direct $CP$ asymmetries for $\ppkstz$ and $\ppkst$ are measured to
be $-0.08 \pm 0.20\pm 0.02$ and $-0.01 \pm 0.19\pm 0.02$, respectively. 
With improved experimental accuracy, the relationships
$\mathcal{B}(\bp \to \ppk) > \mathcal{B}(\bp \to \ppkst)$ and
$\mathcal{B}(\bp \to \ppkst) > \mathcal{B}(\bz \to \ppkstz)$ are established.

We thank the KEKB group for excellent operation of the
accelerator, the KEK cryogenics group for efficient solenoid
operations, and the KEK computer group and
the NII for valuable computing and Super-SINET network
support.  We acknowledge support from MEXT and JSPS (Japan);
ARC and DEST (Australia); NSFC (China); 
DST (India); MOEHRD, KOSEF and KRF (Korea); 
KBN (Poland); MES and RFAAE (Russia); ARRS (Slovenia); SNSF (Switzerland); 
NSC and MOE (Taiwan); and DOE (USA).




\begin{thebibliography}{99}
\bibitem{conjugate} {Throughout this report,
inclusion of charge-conjugate mode is always implied unless
otherwise stated.}
\end{thebibliography}
\end{document}